\documentclass[12pt,a4paper]{article}
 \usepackage{amsmath,amstext,amssymb,amscd}

\begin{document}
\title{On the existence of stable contract systems}
\author{Danilov V.I. \thanks{Central Institute of Economics and Mathematics of the RAS. email: vdanilov43@mail.ru} }
\date{18.05.2025}
\maketitle

In 1962, Gale and Shapley \cite{GS} introduced the concept of stable marriages and proved their existence. Since then, the statement of the stability problem has been highly generalized. And a lot of proofs has emerged for the existence in these more general statements. It's time to review them and identify the similarities and differences.

First, we will briefly discuss the classical case, because the existence proofs in the general case grew out of it.  Or rather, from the idea of "deferred acceptance". When the best of the proposed contracts is temporarily retained until a better offer is received.

\section{The classical case}

Let $G=(V,E)$ be a undirected (multi)graph with vertices $V$ and edges $E$. $E(v)$ denotes the set of edges adjacent to a vertex $v$. Vertices are understood as agents, and edges are understood as possible contracts between the corresponding agents. Agent preferences are set by linear orders $\le_v$ on the set $E(v)$. A subset $S\subseteq E$ is understood as a system of contracts, $S(v)=S\cap E(v)$. A system $S$ is called \emph{stable} if no agent can improve its position by concluding a new contract. More precisely, for any agent, the set of $S(v)$ consists of the best element in $S(v)$, and there are no blocking contracts. A contract $e$ (between $v$ and $w$) is called \emph{blocking} the system $S$ if $e$ is strictly better than $S(v)$ for $v$ and similarly for $w$.

Due to the linearity of the orders $\le_v$, each agent concludes no more than one contract, so any stable system is a matching, so we can talk about stable matchings. It is well known in the matching theory that the case of bipartite graphs is most simply arranged. This is when the set $V$ of vertices is divided into two parts, and edges can only connect vertices from opposite parts. Conventionally, we call the agents of one part \emph{firms}, and of the other part \emph{workers}. In general, stable matchings may not exist. On the contrary, in a bipartite situation, Gale and Shapley showed that stable matchings always exist. In the future, we will consider the bipartite case everywhere, because the general case requires a slightly different approach.

The first proof of the existence of stable systems was given specifically for thus marriage situation. And we want to briefly recall them and trace a development of these ideas to a more general case.

The first proof of the existence of stable matching (and the very concept of stability) was given by Gale and Shapley in \cite{GS}.  They proposed an explicit algorithm (or process) of "deferred acceptance"\!, which led to a very special stable matching. We will not describe their construction in detail due to its well-known nature (see the original work \cite{GS} or the historical account \cite{R}). Let us just remember that this process consists of several consecutive steps. Each step of this process is divided into two stages. At the first stage, "free" workers make offers to firms (from some set available at this point). At the second stage, firms select the best candidate among the received offers (and the worker they have at that moment) and reject all other offers (which reduces the available set of rejected workers). The process continues until "free" workers exhaust their capabilities. The process begins when all the contracts are available to the workers (that is, when the firms have received no offers). And it yields a stable matching.

I so persistently paid attention to the initial state because if steps of the process do not raise objections (and are implicitly present in almost all subsequent proofs), then fixing exactly such an initial state, although it looks natural, is nevertheless not mandatory. The above fixation leads to a distinguished stable matching (the best for workers). However, to get ALL the stable ones, it is worth choosing the initial state less rigidly. We'll come back to this point later.

The above proof is considered as a constructive one. The initial state, the steps of the process that ends in a finite number of steps, and the desired result are clearly indicated. However, due to possible generalizations, "non-constructive" proofs of existence were also of interest. We will discuss two such proofs in more detail. Let us warn right away that the `non-constructiveness' of them is illusory or hidden. To clarify our point, let's imagine that we have a finite poset. It is obvious that there is a maximum element in it, and this statement looks non-constructive, as a pure existence. However, if we turn to the proof, we will see that it is based on the construction of an increasing chain of elements, that is, in fact, it is constructive.

And this is not just an explanatory example. In fact, non-constructive proofs are arranged according to this scheme. A kind of auxiliary poset (or a lattice) is constructed, the maximum elements of which give stable systems. And maximal elements exist according to some well-known theorem. A vivid example of this approach was provided by Adachi \cite{Ad}. He constructed some monotonic mapping of a suitable lattice, the fixed points of which correspond to stable matchings. It remained to use Tarski fixed point theorem. (The fact that he used a lattice and monotonous dynamics, rather than a poset, is not so important. The transitive closure of this dynamic gives a poset. In addition, Tarski theorem also provided a lattice structure on the set of stable matchings, but this is a bit away from the existence problem.)

The Adachi approach was taken up by Fleiner \cite{F}, who applied it in a general situation, which will be discussed below. As an auxiliary set, he considers pairs of subsets in $E$, and defines monotonic dynamics in it, fixed points of which give stable systems. Some variation of Fleiner method is given by Danilov and Koshevoy \cite{DK}. They introduce the concept of \emph{semistable pair} and show that a maximal (relative to some order) semistable pair gives a stable system of contracts.

However, even earlier, before Adachi, a unconstructive proof of the existence of a stable marriage was proposed by Marilda Sotomayor \cite{S}. The idea was to introduce the tricky concept of \emph{quasi-stable} matching. The difference from the stable one was that some blockings were allowed. Namely, it was allowed to block a pair $(f,w)$ if the worker $w$ was unemployed in this matching. Such quasi-stable matchings exist for trivial reasons; for example, empty matching is quasi-stable. An order for such matchings was introduced based on the preferences of the firms. And if we take the maximal quasi-stable matching, then it turns out to be stable. (A similar argument is provided by Schrijver in his book \cite{Sc}.)

The idea of this unexpected trick can be understood with the help of the following `inductive' proof. I came up with it myself, but then I discovered it in \cite{Dw}. Let us select some worker $w_0$ and temporarily remove him. By induction in the resulting "stripped down" problem (there are fewer participants), there is some stable matching $S$. We will return back $w_0$ along with a bouquet of his contracts. Some of them may be of interest to firms. If there are none, then the old matching $S$ remains stable. If there is, then $w_0$ chooses among them the best contract $e_1$ with some firm $f_1$. If this firm was `free', then the new matching $S\cup e_1$ is stable. Otherwise, this firm had a contract $s_1$ with some worker $w_1$, whom it dismisses. If we remove this worker $w_1$, the matching $S-\{s_1\}\cup \{e_1\}$ is stable. We get into a situation similar to the original one and can repeat the previous steps, but with the working $w_1$. It would seem, what have we achieved? The fact that the welfare of the firm $f_1$ has improved (and of the rest firms has not changed). So after a finite number of steps, the process will stop and give a stable matching.

Sotomayor avoids the process by addressing its outcome, maximizing the position of firms. As she explains, `The practical advantage of being nonconstructive is that it gives insights to be used in more complex models, just as happens with proofs of other results valid for this market which do not make use of the algorithm." However, we had to wait almost 30 years for this advantage to be realized.

\section{The general case}

The classical formulation (still bipartite) was gradually generalized. The generalization allows the agents to enter into several contracts (with one or more partners). This immediately raised the question of how to set agent preferences. Eventually, these preferences began to be formulated in the language of choice functions.

A \emph{choice function} (on an abstract set $X$) is a mapping $C:2^X\to 2^X$ such that $C(A)\subseteq A$ for any $A\subseteq X$. It is assumed that an agent with such a choice function, having access to a `menu' $A$, chooses the subset $C(A)$ of `best' (by definition) elements in it. Of course, in order for this choice to be interpreted as a rational choice of the best, certain conditions must be imposed on $C$. In the future, two such rationality conditions will be assumed, namely, consistency and substitutability.

1. \emph{Consistency.} If $C(A)\subseteq B\subseteq A$, then $C(A)=C(B)$.

2. \emph{Substitutability.} If $A\subseteq B$, then $C(B)\cap A\subseteq C(A)$.\medskip

A choice function (further -- CF) with such properties will be called \emph{Plott function}, since it satisfies the so-called `path independence' condition introduced by Plot in \cite{P}
\begin{equation}\label{Plott}
C(A\cup B)=C(C(A)\cup B).
\end{equation}
The proof of this property, as well as other facts about Plott functions, can be found in \cite{matpl}.

These requirements can be explained by different reasons. One of them is that Plott functions generalize linear (and any partial) orders. More precisely, if $\le$ is a partial order on a (finite) set $X$, then the choice of maximal elements from any $A\subseteq X$ is a Plott function. Although Plott functions are not limited by them. Generally, Plott functions are a very successful generalization of the concept of partial order. Another important reason for the interest in Plott functions is that they guarantee the existence of stable systems.

Let us suppose that, for each agent $v$, a Plott function $C_v$ is specified on the set $E(v)$. We still assume that all agents are divided into two groups, firms and workers. A contract system $S$ is  \emph{stable} if it is

(i) \emph{Acceptable}: $C_v(S(v))=S(v)$ for any agent $v$.

(ii) \emph{Unblocked}: if $e$ is a contract between $f$ and $w$, and $e\in C_f(S(f)\cup e)$ and $e\in C_w(S(w)\cup e)$, then $e\in S$.

Here and further we write $A\cup e$ instead of the more correct $A\cup\{e\}$. The first condition is clear - no one wants to abandon the proposed contracts. The second one means that any contract desired by both parties has already been concluded. Or -- if it is not concluded, it "blocks" the proposed system $S$.

One of the advantages of the hypothesis on Plott functions is that we can assume that we have only two agents: Firm and Worker (see \cite{F}). Their CFs are denoted as $F$ and $W$; both are defined on the same set $E$. The definition of stability is also slightly simplified.\medskip

\textbf{ Definition. } A system $S\subseteq E$ is called \emph{stable} if two requirements are met:

           S1. $F(S)=S$, $W(S)=S$.

S2. If a contract $e\in E$ belongs to $F(S\cup e$) and $W(S\cup e)$, then $e\in S$.\medskip

In this situation, stable systems always exist (further, for simplicity, we assume the finiteness of the set $E$). We will show how the proofs outlined in Section 1 are transferred to a more general case. However, it is better to reformulate stability in terms of "desirability", to which we will devote the next section.

Finally, in a situation slightly more general than above, Alkan and Gale \cite{AG} used a process similar to the original Gale-Shapley algorithm to prove the existence. Below we will explain how to turn it into a `non-constructive' proof, replacing the process with its final state. But first it will be convenient to give some facts and concepts about the Plott function.

\section{A digression about desirability}

When working with Plott functions and, in particular, when considering the stability problem, it is convenient to introduce the concept of desirability.

Let $C$ be a Plott function on a set $X$ and $A\subseteq X$. An element $x\in X$ is  \emph{desirable in the state} $A$ if $x\in C(A\cup x)$. For example, $x$ is desirable if $x\in C(A)$. The set of desirable elements is denoted by $D_C(A)$ or $D(A)$.\medskip

\textbf{Lemma 1.} $D_\mathcal C(A)\cap A=\mathcal C (A)$. $\Box$ \medskip

\textbf{Consequence.} $A=\mathcal C(A)$ if and only if $A\subseteq D_\mathcal C(A)$.\medskip

So the operator $D_C$ defines $C$. And it is useful to understand what properties such operators have. The first property is \emph{antimonotonicity}.\medskip

\textbf{Lemma 2.} 1) If $A\subseteq B$, then $D_C(B)\subseteq D_C(A)$.

    2) $D_C(A)=D_C(CA)$.\medskip

Proof. 1) follows from the substitutability. 2) can be seen from Plott relation (\ref{Plott}).

The second property looks like this:
                                  \begin{equation}\label{Lob}
                                  D(A)=D(A\cap D(A))
.                                 \end{equation}
Indeed, $C(A)=A\cap D(A)$, and by Lemma 2, $D(A)=D(CA)=D(A\cap DA)$. This relation is similar to the L\"{o}b identity from the theory of provability \cite{BG}.

I claim that these two properties characterize the operators $D_C$. Namely, let us define a choice function $C=C_D$ using the operator $D$ with these two properties (antimonotonicity and L\"{o}b identity), setting
                                                   $$
                                   C(A)=A\cap D(A).
                                                     $$
It is argued that $C$ is a Plott function.

\emph{Substitutability.} Let $A\subseteq B$. We need to check that $C(B)\cap A\subseteq C(A)$. On the left is $A\cap (B\cap D(B))=A\cap D(B)$. And because of the antimonotonicity, $D(B)\subseteq  D(A)$. So $A\cap D(B)\subseteq A\cap D(A)=C(A)$.

\emph{Consistency.} Let $CA\subseteq B\subseteq A$. From the antimonotonicity, $D(A)\subseteq D(B)\subseteq D(C(A))=D(A\cap D(A))= D(A)$ (due to (\ref{Lob})), so that $DA=DB$. From here
               $$
                               CB=B\cap DB=B\cap DA\subseteq A\cap DA=CA.
                 $$
Since $B\subseteq A$, then from the already proven substitutability $CA=CA\cap B\subseteq CB$. So $CA=CB$. $\Box$\medskip

\section{Existence with two agents}

Let us return to the existence of stable systems in a situation with two agents, Firm and Worker, equipped with choice functions $F$ and $W$ on a set of contracts $E$. From the definition of stability, it is obvious that the system of contracts $S\subseteq E$ is stable if and only if
                                 \begin{equation}\label{S}
                          S=D_F(S)\cap D_W(S).
                                 \end{equation}
An  asymmetric formulation of stability is more interesting.\medskip

\textbf{Proposition 1.} \emph{A contract system $S$ is stable if and only if $S=WD_F(S)$.}\medskip

Proof. For short, we denote $B=D_F(S)$. Suppose that $S$ is stable. By virtue of (\ref{S}) $S\subseteq B$. Therefore (anti-monotonicity) $D_W(B)\subseteq D_W(S)$. According to Lemma 1,
$$
\mathcal{W}B=B\cap D_WB\subseteq D_FS\cap D_WS=S.
$$
(The last equality is valid due to (\ref{S}).) Thus, $\mathcal{W}B\subseteq S\subseteq B$, from where, by virtue of consistency, $\mathcal{W}B=\mathcal{W}S=S$.

Now we show the converse statement, again assuming $B=D_F(S)$. Let $S=WB$. Since $WS=WWB=WB=S$, $S$ is an acceptable system for Worker. The acceptability for Firm is seen from the inclusion $S\subseteq B$ and the equality $FS=S\cap D_F(S)=S$. Now let $e$ be an element desirable for both Worker and Firm in the state $S$. From desirability for Firm, $e\in B=D_F(S)$. From desirability for Worker, we have
$$
e\in W(S\cup e)=W(W(B))\cup e)=W (B\cup e)=W(B))=S. \ \ \ \ \Box
$$

We have obtained a description of stable systems as a fixed point. It remains to understand why fixed points exist. This can be done in two `non-constructive' ways.

The first method follows the `Alkan-Gale way'. Recall that they started from a certain set $B\subseteq E$ (the set of contracts available to Worker at that moment), formed $WB$ (Worker's choice), then $FWB$, and finally received a new set
$$B':=B-(WB-FWB)=(B-WB)\cup FWB.$$
The process ended when $B'=B$, or equivalently when $WB=FWB$. Alkan and Gale started the process from $B=E$ and showed that after stabilization, the set $S:=WB=FWB$ is stable. We claim that the latter is true if $B$ is `big enough'.\medskip

\textbf{Definition.} A system $B\subseteq E$ is called \emph{ample} (or $F$-ample) if $D_F(W B) \subseteq B$.\medskip

In other words, everything that Firm Company wants in the state $W B$ (that is, after Worker's proposition) is in $B$. Obviously, $B=E$ is ample, so ample systems always exist. The benefit of the concept of ampleness is shown in the following statement.\medskip

\textbf{Lemma 3.} \emph{If $B$ is ample and $W B=F W B$, then $S=W B$ is stable.}\medskip

Proof. It is clear that $S$ is acceptable for both Worker and Firm. Suppose now that $e$ is desirable (in the state $S$) for both Firm and Worker. Desirability for Firm means that $e\in D_F(S)$, which, by virtue of ampleness, implies that $e\in B$. Desirability for Worker means that $e\in W(S\cup e)=W(W B\cup e)=W(B\cup e)=W B=S$. $\Box$\medskip

Now let $B$ be minimal (by inclusion) among ample systems. We claim that in this case $W B=F W B$. This follows from the key \medskip

\textbf{Lemma 4.} \emph{If $B$ is ample, then $B'=B-(WB-FWB)$ is also ample.}\medskip

Proof. Let us start by checking a weaker inclusion
                                            $$
      D_F(W B)\subseteq      B'.
                                              $$
Suppose that some $x$ from $D_F(W B)$ falls into $WB-FWB$, in particular, $x\in WB$. Since $x$ is desirable for Firm in the state $W B$, $x\in F W B$. The contradiction.

Thus, $D_F(W B)\subseteq B'$. In particular (since $F W B\subseteq D_F(W B)$), $F W B\subseteq B'$. Tautologically, $F W B\subseteq W B$. Applying the substitutability condition to $B'\subseteq B$, we get
                                                    $$
F W B\subseteq W B\cap B'\subseteq W B'.
                                                      $$

Since $D_F(FWB)=D_F(WB)$ (see Lemma 2, 2)), then by virtue of the antimonotonicity
$$
D_F(WB')\subseteq D_F(FWB)=D_F(WB)\subseteq B'. \ \ \ \ \Box
$$

\textbf{Corollary.} \emph{If $B$ is a minimal ample system, then the system $S=WB$ is stable.} $\Box$ \medskip

In fact, this approach allows us to get ANY stable systems. Indeed, let $S$ be a stable system. Set $B=D_F(S)$. In this case, Proposition 1 gives the equality $S=WB$, which in turn gives the equality $B=D_F(WS)$, which trivially implies ampleness of $B$. \medskip

The second proof, which appeared recently, implements Sotomayor's program. Yang in \cite{Y} found a very nontrivial and elegant implementation of her idea.  \medskip

\textbf{Definition.} Let us say that a system $Q\subseteq E$ is \emph{modest} (Yang talks about quasi-stability) if $Q\subseteq W(D_F(Q))$.\medskip

Note that if we replace $\subseteq$ with equality, we get the stability of $Q$ (Proposition 1). So the modesty is some weakening of the stability. Like Sotomayor, modest $Q$ always exists; for example, an empty $Q$. This hints that if a modest system is `big enough', then it is stable.  \medskip

\textbf{Lemma 5.} \emph{Any modest system is acceptable for both Worker and Firm.}\medskip

Proof. Acceptability for Worker follows from the fact that $Q$ is a subset of $W$-acceptable set  $W(D_F(Q))$. As for Firm, $Q\subseteq W (D_F(Q))\subseteq D_F(Q)$, so the acceptability for Firm follows from Corollary of Lemma 1. $\Box$\medskip

\textbf{Lemma 6.} \emph{If $Q$ is a modest system, then the system $Q'=FWD_F(Q)$ is also modest. }

Proof. By Lemma 2 $D_F(Q')=D_F(WD_F(Q))$. Since $Q\subseteq WD_F(Q)$, by virtue of the antimonotonicity we get the inclusion of $D_F(WD_F(Q)\subseteq D_F(Q)$. So $D_F(Q')\subseteq D_F(Q)$. Applying $W$-substitutability to this inclusion, we get
      $$
           W (D_F(Q))\cap D_F(Q')\subseteq W (D_F(Q')).
      $$
And since $Q'\subseteq W (D_F(Q))$ and $Q'\subseteq D_F(Q')$ (consequence of $F$-acceptability of $Q'$), then $Q'\subseteq W (D_F(Q'))$, that is, $Q'$ is modest. $\Box$ \medskip

 \textbf{Theorem.} \emph{Let $Q$ be a modest system with minimal (by inclusion) set $D_F(Q)$. Then the modest (by Lemma 6) system $S=FW(D_F(Q))$ is stable.}\medskip

Proof. According to Lemma 5, $S$ is acceptable for both Worker and Firm.  In Lemma 6, it was shown that $D_F(S)\subseteq D_F(Q)$. From the assumption of minimality of $D_F(Q)$, we obtain the equality $D_F(S)=D_F(Q)$, so that
 $$
 S=FW(D_F(Q))=FW(D_F(S)).
 $$
We assert that $S=W(D_F(S))$, which gives stability to $S$ by virtue of Proposition 1.

In fact, denote $A:= W(D_F(S))$ (so $S=FA$), and let $a$ be an arbitrary element of $A$. In particular, $a\in D_F(S)$, so that $a\in F(S\cup a)$. Consider the inclusions
                                                      $$
S=FA\subseteq S\cup a\subseteq A
                                                        $$
and apply the consistency of CF $F$. We get $a\in F(S\cup a)=FA=S$. $\Box$\medskip

If a system $S$ is stable, then $S=WD_F(S)$ is a modest system, so this construction also yields all stable systems. \medskip

Both of these proofs are `non-constructive', but behind them there are clearly dynamics, the fixed points of which give stable systems. They are hidden in Lemmas 4 and 6. \medskip

\textbf{Conclusion.} We discussed two general approaches to existence: through ample systems (the Alkan-Gale way) and through modest systems (the Sotomayor-Yang way).

A minimal ample system $B$ yields a stable $S=W B=F W B$. And a maximal modest $Q$ is stable. It seems that these two approaches are simply opposite (or complementary) to each other. It really is.\medskip

Let $B$ be ample, that is, $D_F(W B)\subseteq B$. Then we set $Q=F W B$ and get a modest system.
Indeed, from the substitutability of $W$ and the inclusion of $D_F(W B)\subseteq B$, we obtain the inclusion
                      $$
                          W B\cap D_F( W B)\subseteq  W (D_F(WB)).
                        $$
It remains to be noted (see Lemma 1) that the left-hand side is equal to $F W B$, that is, $Q$. And $D_F(W B)=D_F(F W B)=D_F(Q)$. And we get $Q\subseteq D_F(Q)$.\medskip

Conversely, let $Q$ be modest, that is, $Q\subseteq W(D_F(Q))$. Then we set $B=D_F(Q)$ and get an ample $B$. Indeed, $W B=W(D_F(Q))\supseteq Q$, whence (antimonotonicity) $D_F(W B)\subseteq D_F(Q)=B$.

      \end{document}